\documentclass[12pt]{aastex62}

\begin{document}

\parindent=1.5cm

\title{A Study of Three Intrinsically Luminous Long Period Interacting 
Eclipsing Binaries in the Galaxy and the Small Magellanic Cloud}

\author{T. J. Davidge}
\affiliation{Dominion Astrophysical Observatory,
\\Herzberg Astronomy \& Astrophysics Research Center,
\\National Research Council of Canada, 5071 West Saanich Road,
\\Victoria, BC Canada V9E 2E7\\tim.davidge@nrc.ca; tdavidge1450@gmail.com}

\begin{abstract}

	The properties of the eclipsing binaries 
SMC004701.39-732645.8, SMC005609.41-722809.2 and PW Pup are 
examined using archival photometric and spectroscopic observations. These 
systems have orbital periods of hundreds of days, and contain highly evolved 
components with temperature differences of at least $\sim 1000 - 2000K$. 
The SMC systems are among the brightest eclipsing 
binaries in that galaxy at visible wavelengths, and 
show excess thermal emission that appears to be typical for luminous K 
and M giants in that galaxy. A preliminary mass function for SMC004701 
suggests a minimum mass for the secondary of $3.8 \pm 0.8$M$_{\odot}$, whereas 
the PW Pup mass function indicates a minimum mass for the secondary of 
$3.2 \pm 0.3$M$_{\odot}$. Light curve elements are examined for a range of 
mass ratios using observations at visible and mid-infrared (MIR) wavelengths. 
The results suggest that SMC004701 may prove to be a double line 
spectroscopic binary at visible wavelengths. However, with the possible 
exception of emission from a circumstellar disk, infrared 
observations will likely be required to detect light from the 
secondary in the other two systems. All three systems have likely 
experienced mass transfer, and had components with initial 
masses that correspond to B spectral types on the main sequence. 
The light associated with the secondary likely originates from a 
shell or disk that obscures a hotter central star. 

\end{abstract}

\section{INTRODUCTION}

	The majority of interacting binaries in the solar neighborhood have 
orbital periods log(P$_{day}) < 2$. While there 
are obvious exceptions, most of these contain 
low and intermediate mass stars, with mass transfer that was initiated while 
the primary was evolving on the main sequence or during the early stages of 
post-main sequence evolution. The extent of stellar evolution when interactions 
start is important as the degree of internal evolution prior to interactions 
affects how a star responds to mass loss and accretion, which are key 
factors for defining the end stages of system evolution 
\citep[e.g.,][]{kipandwei1967, geetal2020}.

	Interacting systems with periods log(P$_{day} > 2$) are 
relatively rare and likely had interactions that commenced when at least 
one of the components was more evolved than in shorter 
period systems. Such long period interacting systems may 
be among the most luminous objects in a galaxy, with properties that are the 
result of evolution that is markedly different from that of single stars. 
Some of these systems likely also had massive progenitors. The study of massive 
interacting binaries is of interest as they play important roles in a number 
of astrophysical processes, including the re-ionization 
and on-going chemical evolution of the Universe \citep[e.g.,][]{tsaetal2023}. 
Interactions between massive stars can also define the mass distribution 
of compact relativistic remnants \citep[e.g.,][]{schetal2023}. 

	The number of known interacting binaries with long orbital periods in 
the Galaxy is limited. However, the sample of targets can be expanded by 
examining nearby galaxies, where the panoramic view of the stellar 
content increases the chances of finding rare systems. 
As the closest sizeable galaxies, the Magellanic Clouds play a unique 
role in the study and characterization of such rare objects, and so 
have been a favorite target for photometric campaigns to 
identify variable stars \citep[e.g.,][]{payandgap1966, pay1971, udaetal1997}. 
The OGLE \citep[][and references therein]{udaetal2015} 
and MACHO \citep[][and references therein]{alcetal2000} surveys are of 
particular interest as they obtained light curves for a large number of 
variable stars in the Magellanic Clouds over a time span of several years 
using digital detectors. They are thus obvious datasets from which 
interacting eclipsing systems with long periods can be selected for further 
study. Objects in the Magellanic Clouds are interesting comparison objects 
for their Galactic counterparts as they have chemical abundances that 
systematically differ from those in the solar neighborhood. These differences 
affect the nature of interactions and the subsequent evolution of 
interacting binary systems \citep[e.g.,][]{dav1988, tsaetal2023}.

	Three eclipsing systems with periods that exceed 100 days are examined 
in this study. Two of the binaries are in the SMC and were selected from the 
OGLE survey, while the third is in the Galaxy. The SMC systems were selected 
to be in relatively uncrowded fields. All three have 'EB' type light 
curves, indicating that at least one component is tidally distorted. 
The difference in depths of minima also indicate that 
there is a sizeable temperature difference between the components, as might be 
expected if they are at different stages of evolution or if one 
of the components is masked by an opaque disk. 
The systems have spectral types and/or colors that are indicative of a 
primary that has evolved off the main sequence and is now a yellow supergiant.

	SMC004701.39-732645.8 (OGLE-SMC-ECL-1141; hereafter SMC004701) 
is one of the brightest eclipsing binaries in the Magellanic Clouds. With a 
346 day period it is near the long period end of eclipsing systems 
in the OGLE survey. Light curves that span many cycles were 
also obtained from observations recorded as part of the MACHO 
survey \citep{facetal2007}. Inspection of 2MASS \citep[][]{skretal2006} 
images indicate that the system is relatively isolated, reducing the prospect 
of a significant contribution from field stars when compared with 
systems in more crowded environments. Oddly, SMC004701 is classified 
as a Cepheid in the GAIA DR3. 

	There is evidence that the components of SMC004701 have interacted. 
Both the OGLE and MACHO light curves show cycle-to-cycle scatter, which 
can be a signature of mass transfer. The system is also a source of 
H$\alpha$ emission \citep[][]{lin1961}. 
Photometric measurements made as part of the WISE \citep[][]{wrietal2010} 
survey indicate that there is excess mid-infrared (MIR) 
emission, and this emission is discussed further in Section 2. 
The H$\alpha$ and thermal emission are suggestive of circumstellar and/or 
circumsystem material that might result from mass transfer. 

	Various properties of SMC004701 are summarized in Table 1, where 
the majority of entries are from the GAIA DR3 \citep[][]{gai2023}. 
$\overline{E(bp-rp)}$ is the mean E(bp-rp) of sources with 
Gp $\leq 18$ within 60 arcsec of SMC004701. Only 
objects with parallaxes $\leq 0.05$ masec were considered when 
calculating the mean color excess. The difference 
between E(bp--rp) for SMC004701 and $\overline{E(bp-rp)}$ indicates 
that SMC004701 is more highly extincted than other sources that are within an 
arcmin, and such localized extinction is suggestive of a circumsystem 
shell. However, as radial velocity variations are detected (Section 2.2), 
then the optical depth of the shell is such that 
the signal from the primary is not masked 
at visible wavelengths. The effective temperature and 
surface gravity estimates for SMC004701 are 
consistent with a G type giant, indicating that the primary has 
evolved off of the main sequence.

\begin{deluxetable}{lccc}
\tablecaption{Selected System Properties}
\tablehead{Quantity\tablenotemark{a} & SMC004701 & SMC005609 & PW Pup \\}
\startdata
Period\tablenotemark{b} (days) & 346.3 & 455.9 & 158.0 \\
E$_0$\tablenotemark{b} & 2457339.7301 & 2457166.5385 & 2445340 \\
G (mag) & 15.0 & 14.1 & 9.1 \\
bp-rp (mag) & 1.5 & 1.4 & 1.3 \\
E(bp-rp) (mag) & $0.564^{+0.1}_{-0.1}$ & --\tablenotemark{c} & $0.750^{+0.11}_{-0.37}$ \\
$\overline{E(bp-rp)}$\tablenotemark{d} (mag) & $0.229 \pm 0.035$ & $0.157 \pm 0.027$ & $0.996 \pm 0.081$ \\
$\pi$\tablenotemark{e} (masec) & $0.00 \pm 0.02$ & $0.00 \pm 0.014$ & $0.24 \pm 0.02$ \\
RUWE & 0.96 & 1.04 & 1.74 \\
T$_{eff}$ (K) & $5041_{-21}^{+20}$ & 4355 & $6291_{-123}^{+130}$ \\
log(g) & $2.89_{-0.05}^{+0.02}$ & --\tablenotemark{c} & $1.68_{-0.03}^{+0.01}$ \\
\enddata
\tablenotetext{a}{Most photometric, astrometric, T$_{eff}$, and log(g) entries 
are based on data in the GAIA DR3, except for the effective temperature of 
SMC005609. which is from \citet{kel1999}.}
\tablenotetext{b}{Periods and times of minima from \citet{pawetal2016} and \citet{sametal2017}}
\tablenotetext{c}{Not listed in the GAIA archive.}
\tablenotetext{d}{Mean E(bp-rp) from nearby stars.}
\tablenotetext{e}{Parallax (mas).}
\end{deluxetable}

	Like SMC004701, SMC005609.41-722809.2 (OGLE-SMC-ECL-3367; hereafter 
SMC005609) is one of the brightest eclipsing binaries in the SMC, and also has 
one of the longest periods. It is located close to the 
open cluster NGC 330, which has an age of $\sim 20 - 45$ Myr 
\citep[][and references therein]{patetal2020}, with a 
possible $\sim 8$ Myr age spread \citep[][]{miletal2018}. 
Assuming an SMC-like metallicity, then the main sequence 
turn-off mass for the cluster is in the range 
$\sim 8 - 11$M$_{\odot}$ \citep[][]{dot2016, choetal2016}. Therefore, 
if SMC005609 is a member of NGC 330 then its components had initial 
masses that correspond to those of main sequence B stars. 

	\citet{schetal2008} examined the $I$ light curve of SMC005609 
as part of a survey of variable stars in NGC 330. The 
analysis was restricted to OGLE II measurements, and they 
noted the very red color of the system when compared with those of 
other variables in their dataset. There is H$\alpha$ emission 
\citep[][]{meyandazz1993} that is variable in nature. 
\citet{maretal2007} find that the optical spectrum is consistent 
with that of a cool star, and suggest that the H$\alpha$ emission forms 
in a circumstellar disk around the secondary or in a circumsystem disk.
As shown in Section 2.1, there is also excess IR emission, consistent 
with the notion that the components have interacted. 
The infrared spectral energy distributions (SEDs) of SMC005609 and 
SMC004701 are similar (Section 2.2).

	Various properties of SMC005609 are 
summarized in Table 1, where the effective temperature 
of the primary is that found by \citet{kel1999} from broad-band infrared 
measurements. $\overline{E(bp-rp)}$ was calculated 
using the same criteria as for SMC004701, and 
is lower than that for SMC004701. While SMC005609 does not have 
a surface gravity entry in the GAIA DR3, the primary is likely 
a moderately red giant/supergiant. The effective temperature in Table 1 
is consistent with an early K spectral type.

	The third system is PW Pup, which is a little-studied Galactic
eclipsing system that has a 158 day period \citep[][]{egg1983}. The primary 
eclipse in the \citet{egg1983} light curve appears to be total, opening the
possibility of placing stricter limits on a photometric mass ratio than is
possible for partially eclipsing systems.
It is also one of the longest period EB type eclipsing systems
in the General Catalogue of Variable Stars \citep[][]{sametal2017}.

	\citet{hum1975} assigns the system an
F2epIab $+$ B8 spectral type, where the F supergiant is the dominant
contributor to the light at blue wavelengths. \citet{hum1975} also notes 
that the lower order Balmer lines have P Cygni profiles, indicating an 
expanding medium. However, the MIR SED of PW Pup 
does not show evidence for the large scale excess thermal
emission that is associated with SMC004701 and SMC005609 (Section 2.2). 

	There are indication that the stars in PW Pup are interacting.
\citet{mer1942} identified the system as a source of H$\alpha$ emission.
\citet{bopetal1991} detect prominent near-UV emission, which is a classic
signature of mass transfer \citep[e.g.,][]{plaandkoc1978}, and note that
PW Pup has 'dramatic' mid-UV characteristics when compared with other
systems that contain a hot and a cool component. Still, the mid-UV spectrum
between 2900 and 3000\AA\ in their Figure 9 shows 
decreasing flux towards shorter wavelengths, with a trend that
is steeper than that seen in the other two systems in their study.

	Various properties of PW Pup are summarized in Table 1.
The calculation of $\overline{E(bp-rp)}$ was restricted to objects
with parallaxes $\leq 0.2$ masec that are within 60 arcsec of the system. 
The bp--rp color excess of PW Pup is smaller than the mean color
excess of nearby sources on the sky. The GAIA renomalized unit weight
error (RUWE) of PW Pup in Table 1 is of particular interest as it exceeds the
$\sim 1.25$ threshold suggested by \citet{penetal2022} for stability in the
astrometric parameters. A large RUWE could result if the orbital motions
of the components are discernable by GAIA and/or
if the binary system has an unresolved companion, about which
the PW Pup binary system moves. Evidence is presented in Section 5 
that the faint B star signatures in the \citet{hum1975} spectrum originate 
from a source that is distinct from the stars that make up the 
eclipsing system.

	Given the high luminosities and moderate temperatures of the 
components in these systems then it is likely that 
they have evolved off of the main sequence. 
In this study, emphasis is placed on modelling the light curves of 
these systems and, for two of them, providing limits on 
component masses using archival spectroscopic 
observations. Light curves at wavelengths $< 1\mu$m 
are at the core of the photometic analysis, although 
observations in the WISE W1 ($\sim 3.4\mu$m) and W2 ($\sim 4.6\mu$m) 
filters obtained during the NEOWISE photometric survey \citep[][]{maietal2011, 
maietal2014} are also modeled. While the MIR 
light curves are more sparsely populated than those at shorter 
wavelengths and have been extracted from images that have poorer 
angular resolution than the visible observations, they 
still have the potential to provide additional information 
about the effective temperature of the secondary and other low temperature 
objects, such as shells and disks. MIR observations may 
also pierce the line-of-sight extinction if there is a circumstellar 
and/or circumsystem disk with moderate optical thickness. 

\section{ANCILLARY INFORMATION}

\subsection{Spectral Energy Distributions}

	While photometric observations at visible and red wavelengths have 
traditionally been the primary means of obtaining 
light curve elements, all three systems also have MIR photometry 
that extends out to $24\mu$m. The MIR observations provide insights 
into the nature of cooler objects in the systems, such as the secondaries, 
and circumsystem $+$ circumstellar material. Some of these observations 
were also recorded over multiple epochs, and so can be used to 
construct MIR light curves.

	SEDs in the $1 - 25\mu$m region are shown in Figure 1. Magnitudes are 
in the AB system, and have been normalized to the mean values in the J, H, and 
K measurements (i.e. between $\sim 1$ and $2.5\mu$m). The NIR measurements 
are from the 2MASS \citep[][]{skretal2006} survey, while the MIR 
measurements are from the WISE \citep[][]{wrietal2010} All-Sky data release
\footnote[1]{https://irsa.ipac.caltech.edu/data/WISE/docs/release/All-Sky/index.html}. 

\begin{figure}
\figurenum{1}
\epsscale{1.0}
\plotone{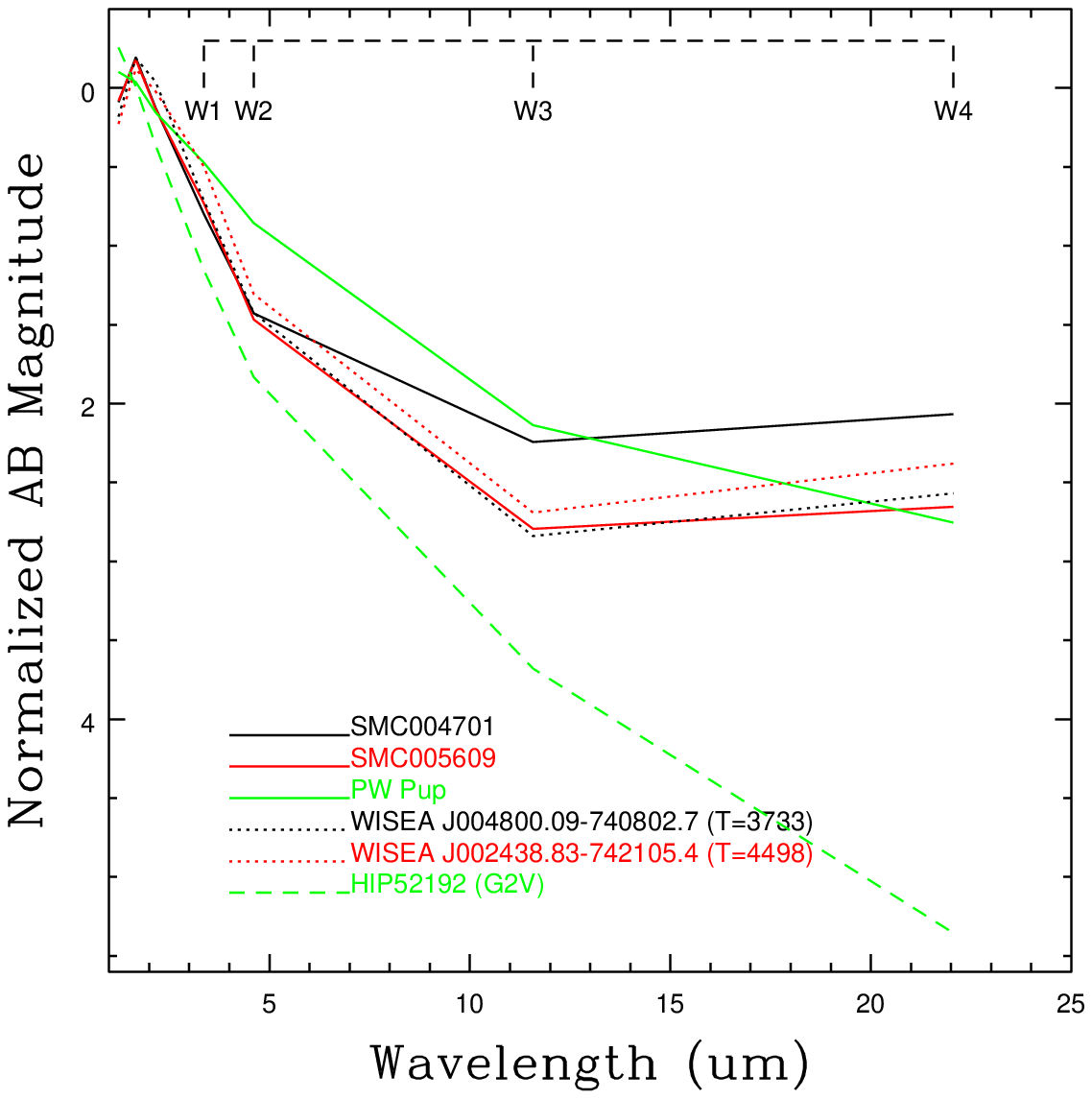}
\caption{Infrared SEDs constructed from 2MASS and 
WISE photometry. AB magnitudes normalized to the 
mean in the $1 - 2.5\mu$m interval are shown. The 
central wavelengths of the four WISE filters are also indicated. The SEDs 
of all three binaries show pronounced MIR emission. However, the differences 
between the shapes of the SEDs of the SMC systems 
(upward breaking for wavelengths longward of $11\mu$m) 
and that of PW Pup (no upward break) is indicative of different 
characteristics of the emitting sources. Also shown are the SEDs (1) for 
two isolated SMC giants that have effective temperatures in the GAIA DR3 
of $3733K$ and $4498K$, and (2) for the G2V star HIP 52192. The latter is 
included to demonstrate the SED of a source that has an intermediate 
temperature with no (or minimal) non-stellar emission. The SEDs of the 
two SMC binaries closely resemble the SEDs of the SMC giants, suggesting 
that the light from the secondaries in these systems may originate 
from circumstellar material associated with those stars that is not the 
result of mass transfer.}
\end{figure}

	The SEDs of two SMC giants that have 
effective temperatures that are comparable to those of the 
secondaries in the SMC systems based on the analysis of the light curves 
are also shown in Figure 1. Aside from having effective temperatures that 
are similar to those of the secondaries, the giants were also selected 
because they are in the low density outer regions of the SMC, and so are 
isolated from stars of similar brightness. The effective temperatures 
provided in Figure 1 for these giants are from the GAIA DR3 archive. 
The SED of the solar neighborhood G2V star HIP 52192 
is also plotted to show the SED of an object that has an effective 
temperature that is not greatly different from those of the primaries, 
but lacks non-stellar MIR emission. 

	There are obvious differences between the SEDs of the binary 
systems. The uptick in the SEDs of the SMC binaries longward 
of $11\mu$m indicates that there is a source (or sources) with 
an effective temperature of a few hundred K. The SEDs of the SMC giants 
also mimic the general shapes of the SMC004701 and SMC005609 SEDs to 
a remarkable degree. As the reference stars are 
presumably not in binary systems that have experienced mass transfer, 
then this suggests that while the MIR emission detected from the SMC binaries 
may be dominated by material associated with the secondary, there is no 
evidence that it originates from an opaque disk that formed as a result of mass 
transfer. In contrast, the MIR emission from PW Pup does not have an upward 
break at longer wavelengths, consistent with the MIR emission originating from 
a warmer source, such as a cool stellar companion without an envelope. 

	Circumstellar and circumsystem material will absorb light. 
E(bp-rp) for SMC004701 exceeds $\overline{E(bp-rp)}$, suggesting 
that the MIR emission and associated extinction is related to the system. 
While there is not an E(bp-rp) estimate for SMC005609 in the GAIA DR3, 
that its SED has a shape that is similar to that of SMC004701 
suggests a similar origin for the MIR emission from that system.

	The reddening estimate for PW Pup in Table 1 
is lower than that of neighboring stars. It is likely 
that extinction is not uniformly distributed along and near this line of sight
given the low Galactic latitude. While the comparatively low 
color excess of PW Pup does not rule out a circumsystem shell, 
the MIR SED indicates that if a shell is present then 
it may be at a larger distance from the system than is the case for the 
SMC systems.

\subsection{Radial Velocities}

	SMC004701 and PW Pup have published velocity measurements that can 
be used to estimate their mass functions, which in turn places lower limits 
on the masses of their secondaries. There are two 
sets of velocity measurements for SMC004701. 
The first of these is a mean radial velocity 
that was found from 8 measurements made by GAIA over a 731.6 day interval. 
A mean of $123.4 \pm 9.4$ km/sec was found; individual velocities are not 
published. The uncertainty in this mean velocity reflects the inherent motions 
of the components as well as the uncertainties in individual measurements.

	An independent velocity measurement of 169 km/sec, with an uncertainty 
of $\leq 5$ km/sec, was made for SMC004701 by \citet{dobetal2014}. This 
differs from the mean GAIA velocity at over the $4\sigma$ level, and 
sets limits to the full amplitude of the radial velocity curve. Arguably 
the most basic approach to place a lower limit on the mass function is to 
simply ignore that the GAIA velocity is a mean that was recorded over a wide 
range of phases, and hence is an estimate of the mean system velocity, 
and instead assume that the GAIA and Dobbie et al. measurements 
are independent and sample different random phases. This yields an 
extreme lower limit to the full amplitude of velocity variations, which is 
ABS($123.4 \pm 9.4$ -- $169 \pm 5$) = $46 \pm 10$ km/sec. The 
minimum half amplitude of the primary star radial velocity 
curve is then $23 \pm 5$ km/sec, and a lower limit for the mass function of 
SMC004701 is thus $0.4 \pm 0.1$ M$_{\odot}$.

	A more realistic estimate of the mass function can be found 
if the orbital eccentricity, period, and time of conjunction are 
known. More specifically, limits can be placed on the amplitude 
of the radial velocity curve of the SMC004701 primary using 
only one velocity measurement as long as the baseline system velocity is known. 
\citet{dobetal2014} observed the portion of the SMC that contains SMC004701 
on the nights of October 18, 20, and 21 2011. Adopting a JD of 2455855 then 
the orbital phase of the observations is $\sim 0.71$, which is fortuitously 
close to an orbital quadrature point -- the peak radial velocity at phase 0.75 
would be $\sim 3\%$ higher than the Dobbie et al. velocity.
While there is a $\pm 2$ day uncertainty in the date 
of observation, this is a minor source of uncertainty when computing the 
orbital phase given the long period of the system. 

	The mean GAIA SMC004701 velocity falls within 
the range measured from field stars for this part of the SMC by 
\citet[][]{harandzar2006} and \citet{dobetal2014}, and so 
the mean GAIA velocity measurement is adopted as the mean system velocity. 
We caution that \citet{harandzar2006} and \citet{dobetal2014} 
find that the velocity dispersion of field stars in the 
region of the SMC that contains SMC004701 is $\sim \pm 25$ km/sec, 
thereby tempering the agreement between the mean GAIA velocity and the 
typical velocity of nearby field stars. This caveat notwithstanding, given that 
the light curve indicates that the orbital 
eccentricity is 0, then the estimated half amplitude of 
the radial velocity curve is $47 \pm 10$ km/sec. 
The resulting mass function is $3.8 \pm 0.8$ M$_{\odot}$, and this is 
adopted as the preferred mass function for SMC004701. 

	A moderately robust estimate of the PW Pup mass 
function can be made using measurements in the GAIA DR3. A radial velocity 
amplitude of 116.2 km/sec is listed for this system, 
based on 13 visits over an 857.2 day interval (roughly 4 orbital cycles). 
An estimate for the maximum radial velocity of the 
primary, as measured with respect to the system velocity, is then 
half of the amplitude, or 58.1 km/sec. The uncertainty in the system 
velocity, which we adopt as the error in the 
mean of the measurements that were used to determine 
the velocity amplitude, is $\pm 11.5$ km/sec.  Assuming that the 
uncertainties in the individual measurements are small when compared 
with the overall velocity variation -- a reasonable assumption 
given the brightness of PW Pup -- then a radial velocity curve for 
the primary with a half amplitude of 58.1 km/sec yields a mass function of 
3.2 M$_{\odot} \pm 0.3$. The mass function may change when a 
more complete set of velocity measurements are obtained. 
Still, it is clear that the secondary has a super-solar mass.

\section{SMC004701}

\subsection{The Light and Color Curves}

	$B, V,$ and $I$ observations of SMC004701 were recorded as part 
of the OGLE III \citep[][]{udaetal2015} and MACHO \citep[][and 
references therein]{alcetal2000} surveys. The most extensive phase 
coverage in the OGLE observations is in $I$, although the 
central regions of secondary eclipse are missed. In contrast, the
MACHO $V$ and $R$ light curves cover all orbital phases. TESS measurements 
of SMC004701 were not considered as there is spotty 
light curve coverage due to the long period of the system. 
The 21 asec angular sampling of the TESS observations also opens the 
possibility of significant contamination from other SMC sources. 

	The $R$ light curve of SMC004701 is shown in the top panel of 
Figure 2. Primary minimum is broad and deep, indicating that at least one 
component has a size that is a large fraction of the orbital separation. 
There are also cycle-to-cycle photometric variations that are 
most obvious in the lower portions of primary minimum, and these are 
also evident in the OGLE photometry. While not shown 
here, these variations are more noticeable in the $V$ light curve, 
suggesting an origin from a source that is hotter than the primary. Similar 
cycle-to-cycle variations are common in the light curves of systems 
where there is on-going mass transfer \citep[e.g.][]{fre1957, 
gieetal2025}, and so their presence in the 
SMC004701 light curve is a possible signature of interactions. 
There is a modest O'Connell Effect \citep[][]{oco1951}, suggesting a 
non-uniform brightness distribution on at least one of the components. 
Such a feature might also be linked to interactions between components 
\citep[e.g.][]{davandmil1984}.

\begin{figure}
\figurenum{2}
\epsscale{1.0}
\plotone{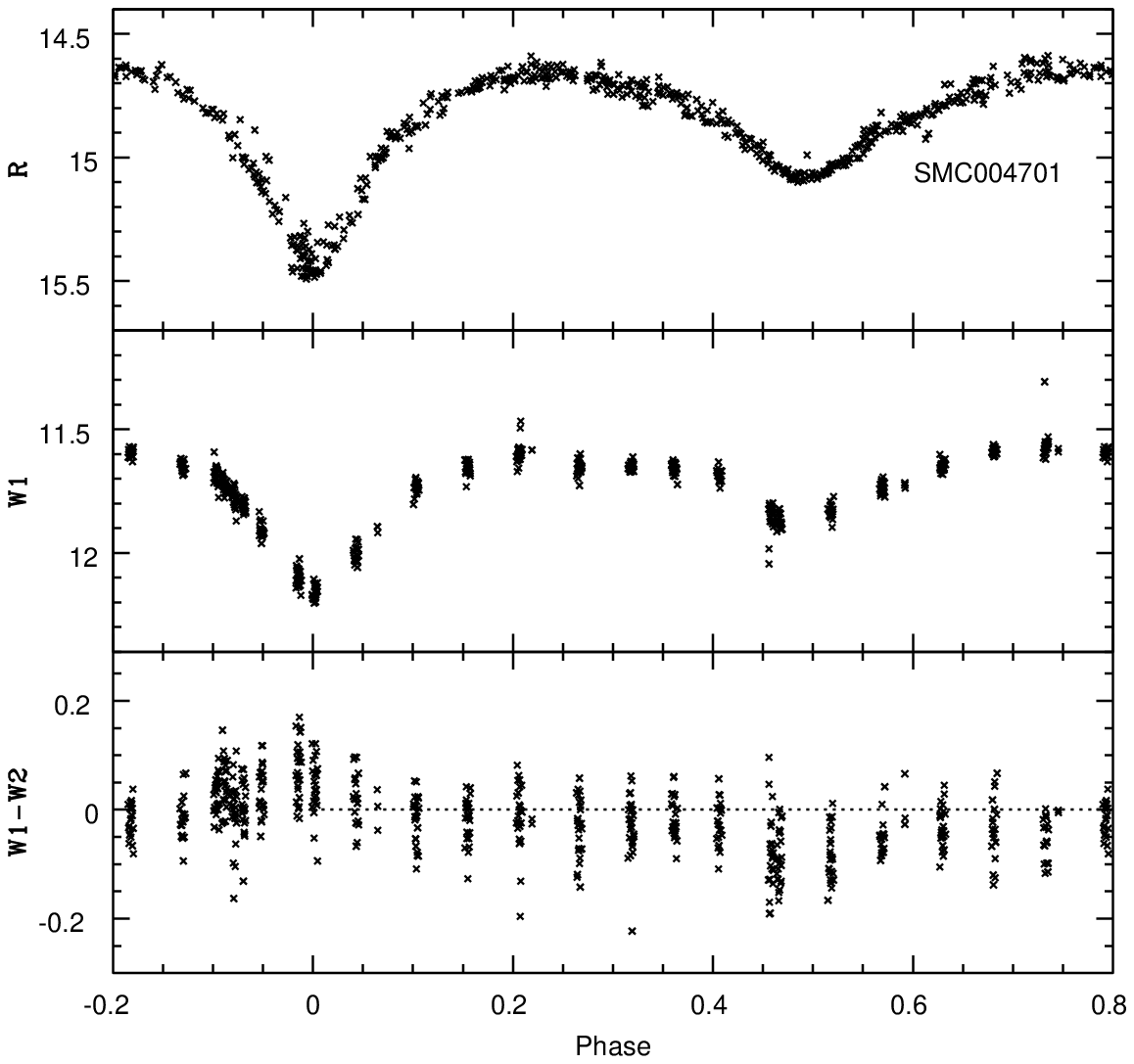}
\caption{Observations of SMC004701 phased according to the \citet{pawetal2016} 
emphemeris. The MACHO $R$ light curve is shown in the top panel, while 
the middle and lower panels show W1 and W1--W2 
measurements. The dispersion in the $R$ light 
curve that is seen in primary minimum and at other phases indicates that 
there are cycle-to-cycle photometric variations similar to those 
associated with interacting systems. In contrast, the scatter 
in the W1 measurements and W1--W2 colors, which were recorded over a number 
of orbital cycles, is comparable to that predicted from the expected 
uncertainties in individual measurements ($\sim \pm 0.02 - 0.3$ mag). The low 
amplitude of these variations in the MIR is consistent with them 
originating from relatively hot sources.}
\end{figure}

	Photometric measurements that were recorded for the NEOWISE survey 
were used to construct the W1 light curve in the middle panel of Figure 2. 
A 0.025 unit shift in orbital phase was required to match 
primary minimum in the W1 and $I$ light curves. 
It is not clear if this shift is related to the cycle-to-cycle variations 
in the properties of the deepest part of primary minimum that are 
seen in the visible light curve. We flag this as a 
photometric characteristic to be examined in future studies of this system 
as longer time baselines are sampled.

	The angular resolution of images recorded in W1 and W2 is $\sim 6$ 
arcsec FWHM \citep[][]{wrietal2010}. While blending is then a concern in the 
SMC, 2MASS and WISE images show that the system is not in a crowded environment 
involving stars of comparable brightness. \citet{dav1987, dav1988} 
examined the environments around early-type eclipsing 
binaries in the SMC and LMC that are not in clusters 
and found that noticeable contaminants typically 
occur at angular offsets in excess of 10 arcsec. 

	The NEOWISE observations were recorded over 10 orbital cycles. 
The spotty phase coverage of the W1 light curve is due to the sampling cadence 
of the NEOWISE survey, coupled with the long period of the system. 
The individual W1 photometric measurements have a predicted uncertainty 
of $\pm 0.02 - 0.03$ magnitudes, and this is roughly consistent 
with the scatter in each concentration of points in the light curve. The 
spread in the W1 light curve at each phase is then dominated by observational 
uncertainties, rather than intrinsic photometric variations, as expected if 
the variations are associated with hot regions. 

	W1--W2 is shown in the bottom panel of Figure 2, 
and the scatter at each phase is consistent with 
that expected from uncertainties in individual measurements. 
W1--W2 is not constant with orbital phase, being 
redder at primary minimum than at secondary minimum. In 
the next section it is shown that the observed variation in W1--W2 is 
larger than expected if it were due solely to stellar eclipses. 
The SED of SMC004701 in Figure 1 suggests that evidence for a 
cool disk may become more apparent in light curves recorded at longer 
wavelengths. Unfortunately, W3 and W4 measurements were not recorded as part 
of the NEOWISE survey, so the phase dependence of the emission 
longward of $\sim 6\mu$m cannot yet be examined.

\subsection{Light Curve Modelling}

	The MACHO $R$ light curve formed the basis for the 
modelling experiments, although the W1 observations were also considered to 
examine the broader applicability of certain system elements. 
Light curve models were compared with unbinned observations in $R$, while 
normal points were calculated for the W1 light curve given the 
phase-clustered nature of those observations. Normal points 
were found by taking the median magnitude in each group of WISE data points, 
which typically included $\sim 20 - 30$ observations that were 
recorded over a two day period. A modest number of normal points are based 
on fewer than 10 observations.

	Model light curves were generated with the PHOEBE 
\citep[][]{prsandzwi2005} package. A metallicity that is appropriate 
for the SMC was adopted. The primary is defined to be the 
hotter star, and its effective temperature, T$_1$, was held 
fixed at the value in Table 1. Limb and gravity darkening 
coefficients were also fixed at standard values based on the properties 
of the components. While there is a modest O'Connell effect, 
spots were not considered. Third light was assumed to be negligible.

	The version of PHOEBE used here did not have provision for 
the W1 and W2 filters. Therefore, the Johnson L and M passbands were 
adopted as proxies for the WISE filters given the similarity 
in central wavelengths. Nevertheless, there is uncertainty 
in the photometric zeropoints of the model magnitudes, and 
so the W1 and W2 photometry was not modelled in an absolute sense. 
Comparisons with the W1--W2 observations are thus differential in nature, with 
a shift applied to match the mean W1--W2 color.

	The difference between the effective temperatures 
of the components indicated by the relative depths of the minima
suggests that the components are likely not in contact, although 
the light curve elements found here suggest that they are close to a 
contact configuration. A semi-detached configuration in which the 
primary is in contact with its Roche lobe was 
adopted for the baseline light curve models. This was a practical choice 
based on the limited information that is currently available for this 
(and the other) systems. While this tacitly imposes 
an evolutionary model, experimentation showed 
that the system elements did not change greatly if other geometric
constraints were imposed, given that both components are close to 
filling their Roche lobes.

	The baseline models do not include an opaque disk around the 
secondary. Such disks are present in some systems that have experienced mass 
transfer, such as $\beta$ Lyrae \citep[e.g.,][]{zhaetal2008} and longer 
period systems like W Cru \citep{pavetal2006}. Disks of this nature have the 
potential to completely obscure the light from the secondary. However, 
there are no spectroscopic observations of SMC004701 (or of the other two 
systems considered here) that could provide information 
about the characteristics of a supposed disk. Thus, 
for SMC004701 we allow T$_2$ to be a free parameter, recognizing that 
it may be the effective temperature of the outer region of an opaque 
structure that encases a hotter object. For the other 
systems, in which there is some evidence for a hotter component, we 
explore models in which the secondary has a high effective temperature, 
but a much lower luminosity than the primary. 

	The similarity between the MIR SED of SMC004701 
and the SMC field giants, which presumably 
are not binaries that have experienced mass transfer, suggests that 
the emission may be intrinsic to the secondary, and 
so need not be a result of mass transfer.
Still, the lack of evidence for an optically thick disk 
notwithstanding, a series of model light curves were generated in which the 
secondary had a distorted shape to simulate the presence of a thick disk. 
The models failed to reproduce the characteristics of the light 
curve, and so disk models were not pursued further. The 
necessity of disk models could be re-visited if and when spectroscopic 
signatures of a disk around the secondary become available.

	As the mass ratio of SMC004701 is not known, 
an initial suite of models that span a range of mass ratios
\footnote[1]{The mass ratio in this study is defined as the ratio of 
the mass of the star eclipsed during secondary minimum (the cooler star) 
with that eclipsed at primary minimum (the hotter star).} were explored to 
compare with the $R$ light curve. The degree of curvature in the 
light curves provides some information about the mass ratio, as 
the brightest component is tidally distorted, 
and so is likely close to filling, or is filling, its Roche lobe. 
Still, mass ratios of 0.7 and 2.0 were the lowest and highest values for 
which a reasonable match could be found to the width and depth of primary 
minimum in the MACHO light curve. Models were 
also generated for a mass ratio of unity to explore an intermediate 
set of elements.

	Differential corrections were applied to $\Omega_2$, 
$i$, and T$_2$ while keeping each mass ratio fixed. All points were assigned 
equal weight. The resulting light curve elements 
are summarized in Table 2, where $1\sigma$ uncertainties are listed. 
The last two rows show the relative bolometric and $V-$band luminosities, 
with the latter based on bolometric corrections from \citet{flo1996}. 
These ratios show considerable model-to-model differences, 
as expected from the range of mass ratios that were explored. They 
suggest that if future studies find that SMC004701 is a double line 
spectroscopic binary in the visible then this will argue for a mass 
ratio $> 1$.

	The uncertainties in each column of 
Table 2 apply only to the mass ratio that was fixed for that 
solution set, and so are extreme lower limits as they do not consider 
the substantial uncertainties in the mass ratio itself. Uncertainties that 
are more realistic can be estimated by comparing the values of T$_2$, $i$, and 
the component radii found for the three mass ratios. This yields 
approximate uncertainties of $\pm 45K$ for T$_2$, $\pm 0.8^o$ for $i$, 
and $\pm 0.05 - 0.06$ for the various stellar size estimates.

\begin{deluxetable}{cccc}
\tablecaption{Light Curve Elements for SMC004701}
\tablehead{Element & q=0.7 & q=1.0 & q=2.0 \\}
\startdata
T$_1$(K) \tablenotemark{a} & 5040 & 5040 & 5040 \\
$\Omega_1$ & 3.24 & 3.75 & 5.25 \\
r$_{1}^{pole}$ & 0.387 & 0.356 & 0.300 \\
r$_{1}^{side}$ & 0.408 & 0.374 & 0.313 \\
r$_{1}^{point}$ & 0.527 & 0.483 & 0.420 \\
r$_{1}^{back}$ & 0.438 & 0.405 & 0.346 \\
 & & & \\
T$_2$(K) & $4332 \pm 19$ & $4422 \pm 6$ & $4346 \pm 8$ \\
T$_2^{W1}$(K) & ($3223 \pm 7$) & ($3051 \pm 55$) & ($3117 \pm 44$) \\
$\Omega_2$ & $2.91 \pm 0.02$ & $3.38 \pm 0.01$ & $4.95 \pm 0.02$ \\
$\Omega_2$(W1) & $3.07 \pm 0.02$ & $3.50 \pm 0.02$ & $5.09 \pm 0.06$ \\
r$_{2}^{pole}$ & $0.383 \pm 0.004$ & $0.408 \pm 0.002$ & $0.441 \pm 0.002$ \\
r$_{2}^{side}$ & $0.414 \pm 0.005$ & $0.440 \pm 0.002$ & $0.473 \pm 0.003$ \\
r$_{2}^{point}$ & $0.611 \pm 0.017$ & $0.609 \pm 0.005$ & $0.611 \pm 0.005$ \\
r$_{2}^{back}$ & $0.515 \pm 0.017$ & 0.$516 \pm 0.005$ & $0.516 \pm 0.005$ \\
 & & & \\
i(deg) & $77.7 \pm 0.3$ & $76.2 \pm 0.2$ & $77.5 \pm 0.1$ \\
i$_{W1}$(deg) & ($70.9 \pm 0.3$) & ($68.5 \pm 1.4$) & ($71.0 \pm 0.8$) \\
 & & & \\
$l^{bol}_1$/$l^{bol}_2$ & 1.67 & 1.13 & 0.81 \\
$l^V_1$/$l^V_2$ & 2.39 & 1.62 & 1.16 \\
\enddata
\tablenotetext{a}{Kept fixed}
\end{deluxetable}

	Model light curves are compared with the observations in the top panel 
of Figure 3. There are departures between the models and the $R$ light curve
for all three mass ratios. The width of primary minimum is slightly narrower 
than predicted by the models, while there are also difficulties matching 
ingress/egress of secondary minimum. We suspect that a spot (or spots) 
may be responsible for some of these differences. 

\begin{figure}
\figurenum{3}
\epsscale{1.0}
\plotone{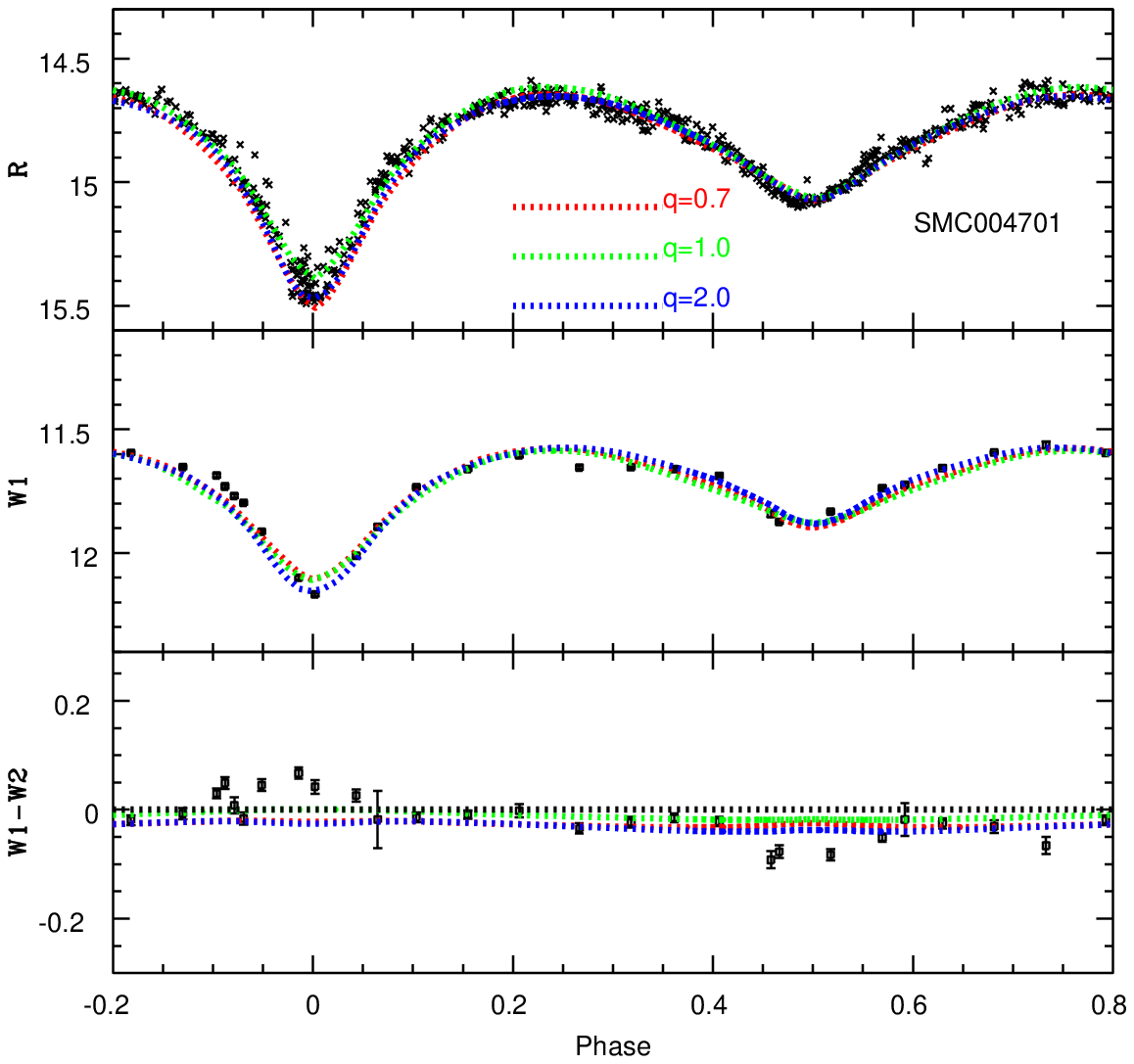}
\caption{Comparing models with the phased MACHO $R$, 
W1, and W1--W2 observations of SMC004701. Normal points are shown 
for the W1 and W2 observations, with errorbars that 
indicate the 1$\sigma$ standard error in the mean. The 
elements used to construct each light curve are listed in Table 2. 
The q=1.0 model departs from the depth of 
primary minimum in $R$, but still falls within the scatter of points 
at this phase. Primary minimum in the W1 light curve is matched 
by all three models. The W1--W2 models underestimate the full 
extent of the color variation, and this is likely due 
to the presence of thermal emission that is more significant 
in W2 than in W1 (e.g. Figure 1).}
\end{figure}

	The W1 observations are of interest given the 
nature of the MIR SED shown in Figure 1. It is then perhaps not surprising 
that model light curves generated from the 
elements based on the $R$ light curve did not match the 
width and depth of the minima in the W1 light curve. 
This is symptomatic of differences in the sources of light that dominate 
the $R$ and W1 light curves.

	The results of applying differential corrections to T$_2$, $\Omega_2$, 
and i based on normal points generated from the W1 observations are shown in 
brackets in Table 2. The orbital inclinations found 
from the W1 light curve are consistently lower than those found in $R$, 
while T$_2$ found from the W1 light curves is systematically lower 
than that estimated from the $R$ light curve (i.e. the W1 light curve 
yields a larger temperature difference between the primary and secondary). 
T$_2$ estimated from the $R$ light curve is consistent with that of a 
K or M type star, while T$_2$ found from the W1 light curve 
is consistent with that of a late M star.

	That T$_2$ and $i$ based on the W1 observations are lower than those 
estimated from the $R$ light curve is likely a consequence of moderately warm 
emission that contributes significantly to the 
MIR light. Evidence to support this claim is presented in the lower 
panel of Figure 3, where it is shown that the model W1--W2 colors, 
which assume that the primary and secondary are the only sources of 
light, yield a smaller color variation than that observed. 
The variation with phase indicates that at least some 
of the thermal emission is likely circumstellar in origin.
The larger than predicted difference in the component 
temperatures occurs because the thermal emission contributes a greater fraction 
of the light in W2 than in W1. Observations 
at longer wavelengths over a range of orbital phases are one 
means of verifying this interpretation. Unfortunately, W3 and W4 
observations of SMC004701 that were recorded for the ALLWISE survey sample 
only four epochs.

	There are broad conclusions that can be drawn from the 
light curve elements listed in Table 2. First, both 
stars are consistently close to -- or are in contact with -- their 
Roche surfaces. This suggests that mass transfer has occurred, or is 
occurring. Moreover, the radius of the secondary is remarkably 
similar over the large range in mass ratio, and indicates that 
this star overflows its Roche lobe in the point direction. In order 
to stay within the Roche lobe then a larger mass ratio approaching 
3 is required. However, if q=3.0 the models predict total eclipses, and 
there are then problems fitting the depth of primary minimum.

	Second, the secondary is $700 - 2000K$ cooler than 
the primary. This confirms its highly evolved nature, indicating that it 
is a cool giant or supergiant, as might be expected 
based on the MIR SED of SMC004701 in Figure 1. As noted previously, it is 
unlikely that all of the MIR light originates from an optically thick disk that 
encases the secondary because the comparison 
of the MIR SEDs in Figure 1 indicates that SMC004701 has similar 
properties to other (presumably single) K and M giants 
in the SMC. Rather, it could be a star that has 
received mass from the primary and is now evolving. If this is the case 
then the mass ratio must be greater than 1.

	The final broad conclusion is that the orbit is inclined to the line 
of sight by between 76 and 78 degrees based on the $R$ light 
curve, and this has implications for determining 
masses from spectroscopic data. he absolute minimum masses for the 
secondary for each mass ratio based on the second (more reliable) 
mass function for SMC004701 found in Section 2.2 are then $\sim 9.2 \pm 
1.9$ M$_{\odot}$ (q=2.0) and $24 \pm 5$ M$_{\odot}$ (q=0.7). The mass 
of the secondary might be even less than the q=2.0 estimate if a slightly 
larger mass ratio is adopted to accomodate the radius of the secondary in the 
point direction. In contrast, the ratio of luminosities listed in Table 2 
will argue for a mass ratio $> 1$ if SMC004701 proves to be a double line 
spectroscopic binary in the visible.

\section{SMC005609}

\subsection{The Light and Color Curves}

	The OGLE $I$ light curve of SMC005609 is shown in the top panel of 
Figure 4. While there are additional OGLE observations in $V$, 
these are limited in number and phase coverage, and so are not considered 
here. As with SMC004701, TESS photometry was not considered due to 
incomplete phase coverage and the crowded nature of the SMC field.

\begin{figure}
\figurenum{4}
\epsscale{1.0}
\plotone{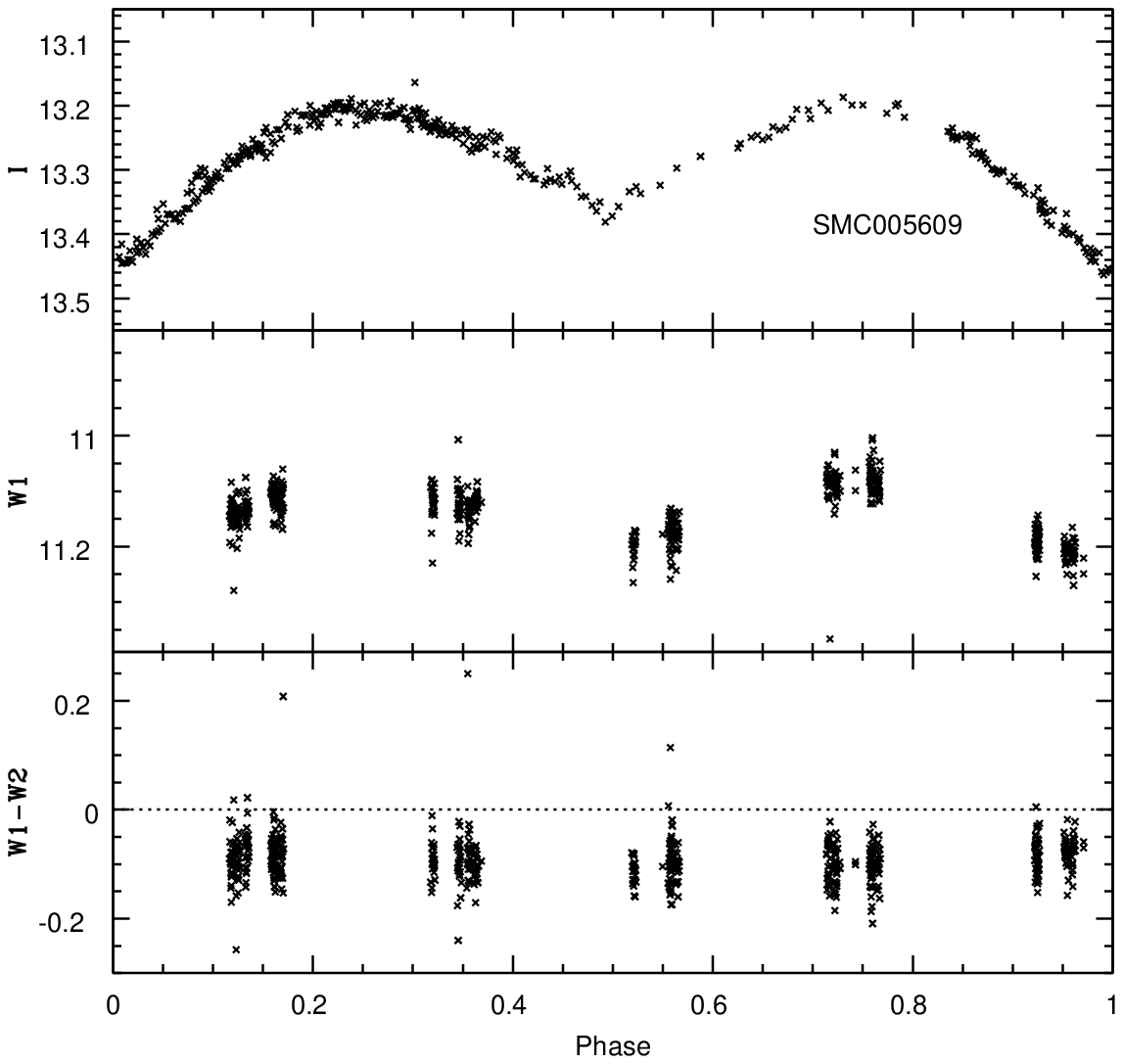}
\caption{Light and color curves of SMC005609, phased 
according to the \citet{pawetal2016} ephemeris. 
The orbital coverage of the WISE photometry is constrained by the long period 
of the system and the sampling strategy of the NEOWISE survey. The scatter 
in the WISE photometry is more-or-less consistent with the predicted 
uncertainties in the individual measurements. There is no evidence 
for transient photometric features and cycle-to-cycle differences in 
the light curves.}
\end{figure}

	The curvature in the light curve suggests that one or both stars are 
tidally distorted and close to filling, if not filling, their Roche 
lobes, and this is consistent with the light curve solutions discussed 
in the next section. The amplitude of the light variations also suggests 
that the angle of inclination may be smaller than for SMC004701. There is only 
modest scatter in the light curve, suggesting that if there is mass transfer 
then it is not inducing transient photometric features in $I$. There is no 
O'Connell Effect. 

	The W1 light curve has spotty phase coverage 
due to an unfortunate combination of the system period and the 
sampling cadence of the NEOWISE survey, which together produced a slow 
year-to-year precession in phase coverage. While the NEOWISE observations 
miss the deepest parts of primary minimum, the coverage of secondary minimum 
is better. W1--W2 appears to be more-or-less constant with orbital phase, 
although there is scatter at each phase point. The scatter in W1 and W1--W2 
is consistent with the uncertainties expected for individual measurements, 
and there is no evidence for cycle-to-cycle variations in the WISE photometry. 

\subsection{Light Curve Modelling}

	Modelling of the SMC005609 $I$ light curve followed the same 
procedures as those employed for SMC004701. With the exception of the models 
that contain a hot dwarf primary that are discussed in the next paragraph, 
a semi-detached configuration in which the primary is in contact with its 
Roche lobe was adopted. A metallicity that is appropriate for the SMC 
was assumed.

	An archival UV STIS spectrum of SMC005609 
\footnote[1]{obs ID: hst\_hsla\_clsngc330arp2-6584, PI: Patrick}, suggests that 
there may be a hot star or other emitting source associated with the system. 
In contrast, \citet{maretal2007} found that the optical spectrum 
of SMC005609 is consistent with that of a cool star, indicating that if a hot 
component is present it does not contribute 
significantly to the visible light. To futher examine 
the influence of a possible faint hot star, an initial series of models 
were explored in which the primary is a B star that is at least $\sim 
0.5$ magnitude fainter than the cooler secondary, which was assigned the 
effective temperature listed in Table 1. The minimum difference 
of 0.5 mag was adopted for these simulations as this is the approximate point 
at which spectroscopic signatures of the fainter star in a system tend to 
become difficult to detect in low to moderate resolution 
optical spectra. As the difference in magnitude
between the components increases then the eclipses of the hotter star become 
progressively weaker, and the variations in the visible light curve are 
dominated by the tidal distortions of the cooler, brighter star. A 
semi-detached configuration was adopted with the secondary in contact with its 
Roche lobe. As noted in Section 3, the object identified as the 
secondary may be a shell around a hotter central star.

	The models with a hot primary could not 
reproduce the basic light curve characteristics, such as the width of 
primary minimum and the shape of the light curve outside of eclipse. 
These experiments indicate that if a B-type 
star is physically close to SMC005609 then it is not part of the eclipsing 
system. All subsequent models were run in which the primary has an effective 
temperature of $4355K$, as estimated by \citet{kel1999}.

	The $I$ light curve of SMC005609 has a smaller amplitude than that 
of SMC004701, suggesting that the system is viewed at an oblique angle. 
This restricts the ability to place limits on possible mass ratios. 
Initial modelling trials involved mass ratios ranging from q=0.25 to 2.0, 
and these confirmed the difficulty in identifying a more restricted 
range of mass ratios for detailed modelling. 

	Selected model light curves are compared with the observations in 
Figure 5, while the light curve elements used to construct these models are 
summarized in Table 3. As with SMC004701, the uncertainties given in 
this table hold only for the mass ratio that was adopted 
for that particular model, and so are lower limits to 
the actual uncertainties in these parameters. Comparing the 
range in parameters shown in Table 3, a more realistic estimated uncertainty in 
T$_2$ is $\pm 200K$, while in $i$ it is $\pm 5^o$. The uncertainties in the 
various size measurements are $\pm \sim 0.10$, due to the wide range in 
mass ratios that were considered.

	The ratio of luminosities in Table 3 suggest that SMC005609 
will be a single line spectroscopic binary at visible wavelengths. 
The orbital inclination is between 55 and $64^o$, 
and so corrections for the orbital inclination when 
determining component masses from radial velocity measurements will be 
significant. While models with q=1.0 and q=2.0 
match the observations, the model with q=0.25 underestimates the 
overall depth of secondary minimum, although there is scatter in the 
photometry near phase 0.5. 

	As noted previously (Section 1), SMC005609 is located close to 
the cluster NGC 330, for which the main sequence turn-off mass is 8 -- 
11 M$_{\odot}$. If SMC005609 is a member of NGC 330 then the initial system 
mass would then have been $\leq 22$M$_{\odot}$. 
The components would then have had intermediate 
initial masses and B spectral types when on the main sequence.

\begin{figure}
\figurenum{5}
\epsscale{1.0}
\plotone{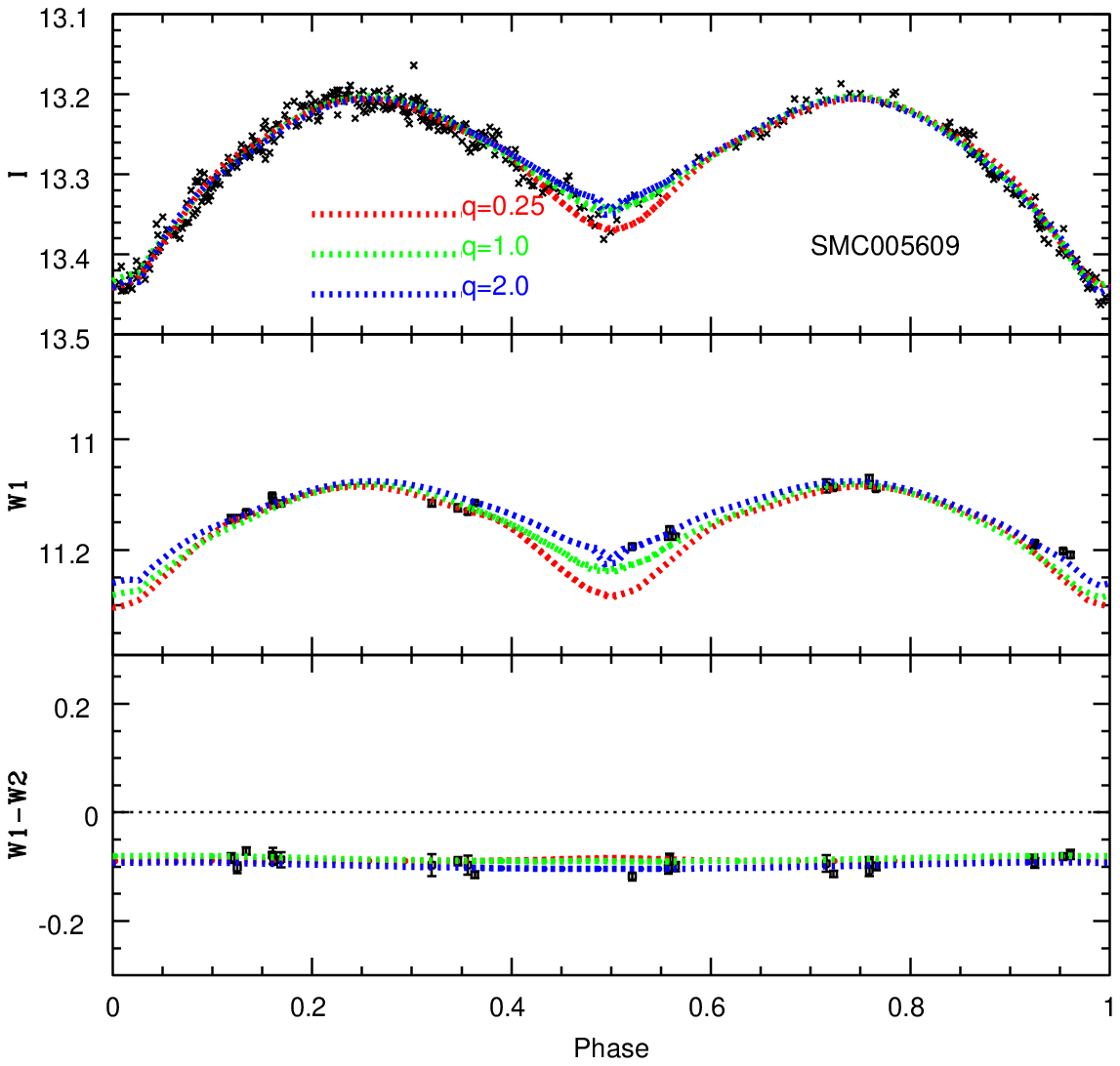}
\caption{Comparing model light and color curves with $I$, 
W1, and W1--W2 observations of SMC005609. The $I$ light curve 
is matched by models that span a wide range of mass ratios 
due to its shallow nature. The W1 and W2 points are normals calculated 
from each 1 - 2 day block of observations, and the errorbars show $1\sigma$ 
errors in the mean. The quantized nature of the W1 observations limits 
the ability to estimate $i$ from those observations alone, and so the models 
in the middle panel adopt the orbital inclinations and temperatures for 
the secondary found from the $I$ light curve. 
The W1 observations near phases 0.55 and 0.95 fall above some of the 
models. The W1--W2 colors are well matched by the models, albeit with 
the caveat of limited phase coverage. There is no evidence of a 
phase dependent thermal component like that seen in the W1--W2 colors of 
SMC004701 in Figure 3.}
\end{figure}

\begin{deluxetable}{cccc}
\tablecaption{Light Curve Elements for SMC005609}
\tablehead{Element & q=0.25 & q=1.0 & q=2.0 \\}
\startdata
T$_1$(K) \tablenotemark{a} & 4355 & 4355 & 4355 \\
$\Omega_1$ \tablenotemark{a} & 2.35 & 3.75 & 5.25 \\
r$_{1}^{pole}$ & 0.470  & 0.356 & 0.300 \\
r$_{1}^{side}$ & 0.508  & 0.374 & 0.313 \\
r$_{1}^{point}$ & 0.627 & 0.489 & 0.420 \\
r$_{1}^{back}$ & 0.536 & 0.405 & 0.345 \\
 & & & \\
T$_2$(K) & $3648 \pm 15$ & $3368 \pm 7$ & $3344 \pm 7$ \\
T$_2^{W1}$(K)\tablenotemark{b} & (--) & (--) & (--) \\
$\Omega_2$ & $2.46 \pm 0.01$ & $3.78 \pm 0.01$ & $5.39 \pm 0.01$ \\
r$_{2}^{pole}$ & $0.225 \pm 0.002$ & $0.353 \pm 0.001$ & $0.403 \pm 0.002$ \\
r$_{2}^{side}$ & $0.232 \pm 0.002$ & $0.370 \pm 0.001$ & $0.426 \pm 0.002$ \\
r$_{2}^{point}$ & $0.264 \pm 0.004$ & $0.458 \pm 0.007$ & $0.496 \pm 0.005$ \\
r$_{2}^{back}$ & $0.251 \pm 0.003$ & $0.399 \pm 0.002$ & $0.450 \pm 0.002$ \\
 & & & \\
i(deg) & $65.2 \pm 0.1$ & $55.0 \pm 0.1$ & $54.0 \pm 0.1$ \\
i$_{W1}$(deg)\tablenotemark{b} & (--) & (--) & (--) \\
 & & & \\
$l^{bol}_1$/$l^{bol}_2$ & 9.20 & 2.88 & 1.62 \\
$l^{V}_1$/$l^{V}_2$\tablenotemark{c} & 49.00 & 15.13 & 8.52 \\
\enddata
\tablenotetext{a}{Kept fixed.}
\tablenotetext{b}{Inclinations and T$_2$ from the $I$ light curve were 
adopted for the W1 models.}
\tablenotetext{c}{Uncertain given the low effective temperature of the 
secondary, and the resulting uncertainties in the bolometric corrections.}
\end{deluxetable}

	Normal points were constructed for each 
concentration of WISE measurements that were recorded over a 1 -- 2 day 
period. The modest amplitude of the light curve at 
MIR wavelengths, combined with the highly quantized phase sampling, 
prevented obtaining independent estimates for T$_2$ and $i$ from the W1 light 
curve. Therefore, W1-based entries for T$_2$ and $i$ are not 
listed in Table 3.

	Some of the models in the middle panel of Figure 5 fail to match 
the W1 light curve near phases 0.55 and 0.95, 
even though the corresponding models more-or-less match the $I$ 
light curve at these phases. Still, the model W1--W2 curves 
track the normal points over a range of phases, suggesting that there 
are not concentrated sources of MIR emission at any point 
in the orbit. This is consistent with the dominant source of MIR emission being 
uniformly distributed along the orbital plane, as might be expected if it is 
circumsystem in origin. This is in contrast to SMC004701, where undulations 
in the W1--W2 curve are indicative of a circumstellar component with a 
non-uniform distribution.

\section{PW Pup}

\subsection{Light and Color Curves}

	As with SMC004701 and SMC005609, there is not a complete 
TESS light curve for PW Pup. However, there are two other 
largely complete photometric datasets. \citet{egg1983} 
published photelectric measurements with the most extensive 
coverage in $V$ \footnote[1]{The entries in Table 1
of \citet{egg1983} are labelled 'V', although the discussion in his text 
suggests that these were made in the y filter. For spectral type G0, the 
difference between $V$ and y magnitudes amounts to a few hundredths of a 
magnitude \citep[][]{str1973}. Observations in a standard V filter 
are presented in Table 2 of \citet{egg1983}, and these agree with 
magnitudes made at similar times in Table 1 at the $\sim \pm 0.02$ magnitude 
level. Therefore, the entries in Table 1 of \citet{egg1983} can be adopted 
as being in $V$.}. Primary minimum in the \cite{egg1983} light curve has 
a more-or-less flat bottom, suggesting a total 
eclipse. \citet{egg1983} also recorded observations in $b, v, 
u$, and H$\beta$. $b-y$ is more-or-less constant with phase. 
While few in number, the H$\beta$ observations vary with orbital phase, 
with the weakest measurement within primary minimum. An O'Connell Effect is not 
present, although the phase coverage near the quadrature points has gaps.

	A second dataset was obtained in $V$ as part of the All Sky 
Automated Survey (ASAS; \citet{poj1997, poj2002}). These data 
were recorded over $\sim 3500$ days ($\sim 22$ orbital cycles), and 
reveal cycle-to-cycle variations in the light curve. 
These variations are present in primary minimum, which has a flat bottom at 
some epochs, similar to the \cite{egg1983} light curve, and a pointy 
bottom at other epochs, suggesting a partial eclipse. Given these 
variations and the large number of cycles covered then the 
ASAS observations were adopted for light curve modelling. The 
ASAS light curve is shown in the top panel of Figure 6. The measurements 
used here are those made in aperture 2 and were restricted to those 
classified as either grade A or B.

\begin{figure}
\figurenum{6}
\epsscale{1.0}
\plotone{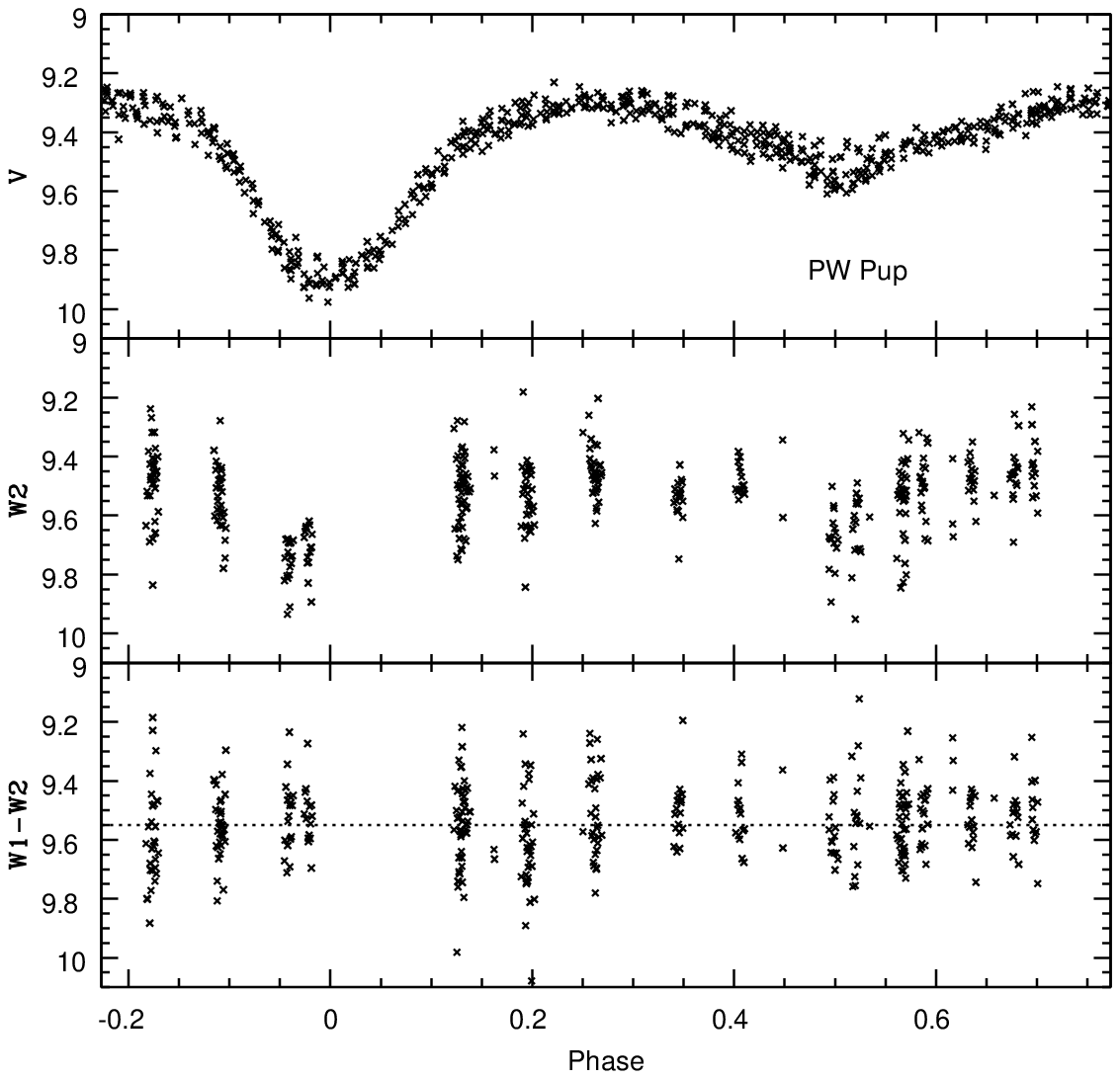}
\caption{Light and color curves of PW Pup. The ASAS $V$ measurements are 
those made in aperture 2 with a quality rating of either grade A 
or B. The W2 light curve, shown in the middel panel, is adopted for 
modelling as observations in that filter are less susceptible to saturation 
than those in W1. In fact, median normal points generated from the W2 
observations track models based on elements found in $V$ (Figure 7). 
The scatter in W1--W2 is likely a consequence of saturation in W1. The phase 
coverage of the WISE photometry is spotty due to the long period 
of the system and the sampling cadence of the NEOWISE survey.} 
\end{figure}

	The W2 light curve and W1--W2 colors of PW Pup are shown 
in the middle and lower panels of Figure 6. 
PW Pup is $3 - 4$ magnitudes brighter than SMC004701 and SMC005609, 
and the W1 observations are susceptible to saturation. 
Therefore, the W2 light curve was adopted for modelling 
as measurements in that filter are less susceptible to saturation. 
An inspection of 2MASS and WISE images indicates that 
crowding is likely not an issue for the WISE photometry, although the 
large RUWE entry in Table 1 could suggest the presence of a third body 
on sub-arcsec scales.

	As with the SMC systems, the WISE light curve of PW Pup has spotty 
phase coverage due to the long orbital period and the 
sampling cadence of the NEOWISE survey. The substantial dispersion in 
W2 and W1--W2 is likely an artifact of outliers due to 
target saturation rather than intrinsic photometric variations. 
Primary and secondary minima are clearly evident, and it is 
demonstrated below that these have depths that are consistent with 
the $V$ light curve elements. 

\subsection{Light Curve Modelling}

	Procedures similar to those described for SMC004701 and SMC005609
were followed to model the PW Pup light curves. 
As with SMC005609, there are possible signatures from a faint B star in the 
spectrum of PW Pup \citep[][]{hum1975}, and one possible interpretation 
is that such a star could be the primary. To better understand the role that 
a faint B star may play in the system, an initial set of models were
run in which the primary is a late B star and the secondary is an F
supergiant that is in contact with its Roche lobe. As was done for 
SMC005609, a series of models were generated in 
which the B star was at least 0.5 magnitudes
fainter than the supergiant, in keeping with its modest contribution
to the blue spectrum of the system. As was the case for SMC005609, 
the resulting models could not reproduce the amplitude 
or eclipse properties of the light curve. Therefore, we conclude 
that the B star is not a member of the eclipsing 
system, but could still be an unresolved close companion. T$_1$ was then 
fixed at the value in Table 1 for subsequent models, while T$_2$ was left 
as a free parameter to be determined from the light curve models. A 
solar abundance was assumed. 

	The light curve elements shown in Table 4 assume 
a semi-detached configuration with the primary in contact with its Roche lobe. 
Model light curves are compared with the observations in Figure 7, and a 
flat bottom for primary minimum is not matched by any of the models. 
There is also difficulty matching the width of primary minimum, and this 
is reflected in the physical size of the secondary, 
which differential corrections drive to slightly overflow its Roche lobe 
for all three mass ratios. The width of primary 
minimum is most closely matched by the q=5.0 model, although the 
relative size of the cool secondary with respect to the primary in that 
model has a major impact on the MIR light curve, and its ability to match the 
observations (see below). There is a range of $\sim 200K$ between the 
T$_2$ estimates, and these suggest that the secondary is then a late G 
or early K giant. The ratio of luminosities in Table 4 suggest that 
PW Pup will be a single line spectroscopic binary at visible 
wavelengths unless q $\sim 5$.

\begin{deluxetable}{cccc}
\tablecaption{Light Curve Elements for PW Pup}
\tablehead{Quantity & q=0.40 & q=1.0 & q=5.0 \\}
\startdata
T$_1$(K) \tablenotemark{a} & 6300 & 6300 & 6300 \\
$\Omega_1$ & 2.68 & 2.88 & 9.16 \\
r$_{1}^{pole}$ & 0.432 & 0.414 & 0.233 \\
r$_{1}^{side}$ & 0.461 & 0.439 & 0.242 \\
r$_{1}^{point}$ & 0.580 & 0.552 & 0.330 \\
r$_{1}^{back}$ & 0.488  & 0.467 & 0.274 \\
 & & \\
T$_2$(K) & $4480 \pm 14$ & $4668 \pm 10$ & $4630 \pm 2$ \\
T$_2^{W2}$(K) & (--)\tablenotemark{b} & (--)\tablenotemark{b} & (--)\tablenotemark{b} \\
$\Omega_2$ & $2.59 \pm 0.01$ & $2.78 \pm 0.01$ & $9.12 \pm 0.01$ \\
r$_{2}^{pole}$ & $0.300 \pm 0.002$ & $0.317 \pm 0.002$ & $0.489 \pm 0.001$ \\
r$_{2}^{side}$ & 0.$316 \pm 0.003$ & $0.334 \pm 0.002$ & $0.533 \pm 0.002$ \\
r$_{2}^{point}$ & $0.528 \pm 0.005$ & $0.564 \pm 0.004$ & $0.683 \pm 0.001$ \\
r$_{2}^{back}$ & $0.363 \pm 0.005$ & $0.380 \pm 0.004$ & $0.556 \pm 0.001$ \\
 & & \\
i(deg) & $78.8 \pm 0.5$ & $76.8 \pm 0.2$ & $66.7 \pm 0.1$ \\
i$^{W2}$(deg) & (--) & (--)\tablenotemark{a} & ($75 \pm 1$)\tablenotemark{c} \\
 & & & \\
$l^{bol}_1$/$l^{bol}_2$ & 7.70 & 5.41 & 0.77 \\
$l^{V}_1$/$l^{V}_2$ & 12.51 & 8.80 & 1.26 \\
\enddata
\tablenotetext{a}{Kept fixed.}
\tablenotetext{b}{Not changed, as the elements from the $V$ light curve 
produced a match to the W2 light curve.}
\tablenotetext{c}{Uncertainty estimated from light curve fitting experiments.}
\end{deluxetable}

\begin{figure}
\figurenum{7}
\epsscale{1.0}
\plotone{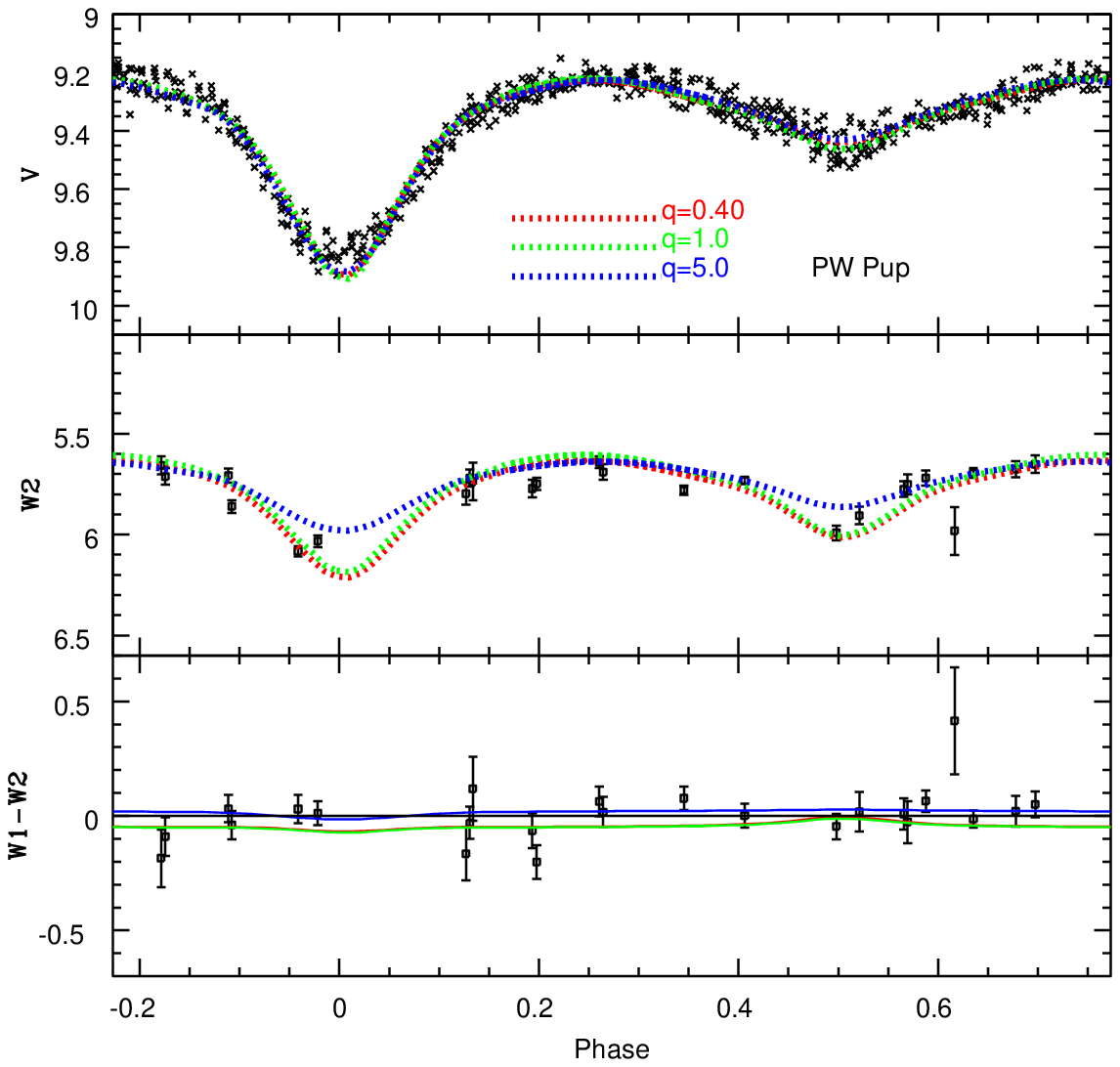}
\caption{Comparing model light curves with the $V$, 
W2, and W1--W2 observations of PW Pup. Normal points are 
plotted for W2 and W1--W2, with the error bars showing $1\sigma$ 
uncertainties about the mean. The q=5.0 model is based on the 
elements found from the $V$ light curve, and it does 
not match the depth of the minima in the W2 observations. An 
inclination near 75$^o$, which is $\sim 8$ degrees larger than that found 
at visible wavelengths with the same mass ratio would produce a model that 
better matches the W2 light curve. The W2 observations near phase 
0.57 depart significantly from the q=0.4 and q=1.0 models. 
All three models more-or-less match the trend 
defined by the W1--W2 normal points at most phases. While the 
q=5.0 model colors are offset from those of the q=0.4 and q=1.0 models, 
they still fall within the observed color envelope. 
Points with the reddest W1--W2 color are concentrated 
between phases 0.2 and 0.35, suggesting a source of warm emission 
between the components that is visible at these phases.}
\end{figure}

	The W2 light curve of PW Pup can be matched 
with the elements obtained from the $V$ light curve for mass 
ratios 0.4 and 1.0. However, the situation is different if q = 5.0, as 
models based on the $V$ light curve elements do not match the depths 
of the minima. This discrepancy can be rectified if an orbital 
inclination of 75$^o$ (i.e. $\sim 8$ degrees larger 
than that found in $V$ for the same mass ratio) 
is adopted. The inability to match the light curve with elements found 
in $V$ for q=5.0 is due to the comparatively large size of the cool 
secondary with respect to the hotter primary at this mass ratio, which 
reduces the depths of eclipses when compared with models that have 
smaller mass ratios, and hence components with more similar sizes.

	There is a slight bump in the W2 light 
curve near phase 0.55 that does not match the models with q=0.4 
and q=1.0. This bump is made up of observations that were recorded over a 
range of epochs, and so it is not a short-term transient feature. 
There is a single point near phase 0.57 in the $V$ light curve 
that also falls above the models, but the significance of this is unclear 
given that it is only a single observation. 
The W1--W2 color for the points that define the bump 
are not markedly different from those at other phases. 

	The model W1--W2 curves have trends that more-or-less match 
the observations, with the q=5.0 model colors offset slightly from those 
based on q=0.4 and q=1.0 models. There are normal points 
between phases 0.2 and 0.3 with small error bars that have some of the 
reddest W1--W2 colors observed over all phases. These points and the bump 
seen near phase 0.55 could suggest the presence of material with an 
effective temperature of $\sim 1000K$ that is visible between the components.
This possibility could be explored with observations at wavelengths 
longward of $6\mu$m.

	As expected given the large spread in mass ratios that have been 
examined, there is a huge range of potential component masses. 
The mass function calculated in Section 2.2 from the amplitude 
of the velocity measurements in the GAIA DR3 yields a mass for the secondary 
of $41.5 \pm 0.4$M$_{\odot}$ if q=0.4, while the 
primary would then have a mass in excess of 100M$_{\odot}$. In 
contrast, with q=5.0 then the secondary has a mass of $5.9 \pm 0.5$M$_{\odot}$, 
and the primary has a mass of only $1.2 \pm 0.1$M$_{\odot}$. 

	We caution that the large RUWE in Table 1 
suggests that another component may be present, 
and the B star that was detected by \citet{hum1975} 
is a possible candidate for such an object.
If an unresolved companion contributes a significant amount 
of light then this will affect light curve elements, and the ability 
to match the observations. Searching for evidence
for light from a (presumably unresolved) third component near PW Pup is thus
important for understanding the evolutionary state of this system
and estimating component masses. That the q=0.4 and q=1.0 models 
based on the $V$ light curve match the W2 observations suggests 
that if a source of third light is present then it 
cannot be a luminous cool star. 

\subsection{Comparison with Li et al. (1989)}

	The \cite{egg1983} observations were 
examined previously by \citet{lietal1988}, who 
find an inclination of $61.1 \pm 0.6^o$ and a photometic mass ratio of 
$1.56 \pm 0.02$. Based on the mass function found in Section 2.2, then these 
parameters yield a $\sim 13$M$_\odot$ secondary 
and a $\sim 8$M$_\odot$ primary. The \citet{lietal1988} model 
assumes an overcontact configuration, despite the 
$\sim 7000K$ difference in the component temperatures 
that is predicted by their model. While this temperature difference 
greatly exceeds what is found here, it should be recalled that the elements in 
Table 4 also have a secondary that overflows its Roche lobe, while the 
primary is constrained to fill its Roche lobe. In any event, there are 
considerable differences between the Li et al. model light curve and 
the observations. For example, the model does not track the observations 
outside of primary eclipse, especially near secondary minimum. 

\section{DISCUSSION AND SUMMARY}

	The light curves of three intrinsically luminous 
long period eclipsing binaries have been examined. SMC004701, SMC005609, and 
PW Pup have not been studied extensively in the past, likely due to 
the difficulties inherent to obtaining complete observational data given their 
long periods and, in the case of the two systems in the SMC, their 
recent discovery. The SMC systems are among the brightest eclipsing 
binaries at visible and red wavelengths yet found in that galaxy. 
Understanding these objects and estimating their masses is thus important for 
exploring the nature of similar systems in other external galaxies. 

	Light curves that sample MIR wavelengths have been examined 
alongside their visible/red counterparts. MIR light curves 
are of interest as they can provide additional information about 
sources in and around the system that may have moderately low temperatures, 
including cool companions and structures related 
to mass transfer, such as disks. The two SMC systems have MIR SEDs that 
are indicative of excess emission at wavelengths $> 5\mu$m, and that 
are similar to the SEDs of (presumably isolated) late-type SMC supergiants. 

	The elements found from MIR light curves may differ from those 
found at shorter wavelengths. For example, lower values for T$_2$ and $i$ 
than those estimated at shorter wavelengths were required 
to match the MIR light curve of SMC004701. This is 
symptomatic of contamination at all phases from a bright MIR component, 
and a circumsystem shell that may have formed during 
mass transfer is one possible source of such emission.
That the W1--W2 color curve for SMC004701 has an amplitude that is greater 
than that predicted by the models suggests that the source 
of at least some thermal emission is not uniformly distributed, with 
some emission originating from a circumstellar shell or mass stream.

	The MIR observations used here have 
comparatively low angular resolutions when compared with the visible/red 
photometry (6 arcsec FWHM $vs$ 1 arcsec FWHM), 
which results in a greater susceptibility to contamination from light 
that does not originate from the eclipsing system. 
In recognition of this the SMC systems were selected because they 
are in relatively uncrowded environments. While SMC005609 is a possible 
member of NGC 330, it is in the outer regions of that cluster.

	The periods and intrinsic brightnesses of 
these systems make them potential candidates for hosting massive stars 
and/or for having components that experienced interactions during more advanced 
phases of stellar evolution than shorter periods systems. 
It is then not surprising that two of the systems (SMC004701 
and PW Pup) have spectral types that indicate 
the primary has evolved off of the main sequence. 
While spectroscopically determined mass ratios are not known for these 
systems, limits can be placed on mass functions for SMC004701 and 
PW Pup using existing data. These indicate that the secondary in SMC004701 
has a minimum mass of $3.8 \pm 0.8$M$_{\odot}$ while the secondary in PW Pup 
has a minimum mass of $3.2 \pm 0.3$M$_{\odot}$.

	Light curve models that assume a semi-detached 
configuration in which the primary is in contact with its Roche 
lobe were considered. This configuration was adopted given the substantial 
differences in the effective temperatures 
of the components and their highly evolved natures. 
In any event, the secondary is found to be 
close to filling its Roche lobe in each system, and experimentation 
found that adopting models in which the secondary is in contact with its Roche 
lobe did not markedly change the light curve elements.

	Mass ratios are a critical piece of information for interpreting the 
evolutionary states of these systems. Unfortunately, estimating 
mass ratios for these systems is problematic at present. Moreover, the 
ratio of component brightnesses based on the light curve elements 
suggests that observations in the IR may be required to detect the 
spectrum of the secondary and monitor its motion.

	Rather than attempt to determine a set of 
definitive elements for each system, model light curves 
were explored for a range of mass ratios to examine the range of 
possible elements for different evolutionary states, and identify 
quantities that have a low sensitivity to the mass ratio. There are only 
modest differences between the three light curves generated for each system, 
even though substantial differences in mass ratio were considered. 
This demonstrates the well-known problematic nature of determining a mass ratio 
from photometric data alone except in special cases.

	The model light curves place broad limits on the 
temperatures of the components and the system 
inclinations. The orbital inclination of SMC004701 falls 
between 76 and 78 degrees, while the temperature of the secondary 
based on the $R$ observations is $4332 - 4422K$. However, while some 
of the system elements are well-defined, only very broad conclusions 
about the initial system mass can be drawn. Adopting the mass function based on 
phased radial velocity measurements in Section 2.2 then the 
mass estimates for the secondary in SMC004701 are $12 \pm 2$ M$_{\odot}$ 
(q=2) and $37 \pm 8$ M$_{\odot}$ (q=0.7). 
The components of SMC004701 then likely had intermediate to moderately 
high initial masses.

	The orbital inclination estimates of SMC005609 fall between 54 and 
65$^o$, while the temperature of the secondary is $3344 - 3648K$. 
Model light curves that have a B star that 
is fainter than the cooler secondary, as suggested by 
optical spectra, cannot reproduce the observed light curve, 
indicating that if a B star is present then it is not 
part of the eclipsing system. Unfortunately, in the absence of velocity 
measurements, no conclusions can be reached about the initial masses of 
the SMC005609 components. Still, this system has a period 
and MIR SED that is similar to those of SMC004701. 
Moreover, if SMC005609 is a member of the cluster NGC 330 then it 
must have intermediate initial masses. 

	For PW Pup the inclination is likely in the range 67 -- 79$^o$, 
while the temperature of the secondary is $4480 - 4668K$. 
The mass estimates for PW Pup are highly uncertain 
because of the broad range of possible mass ratios. Still, the 
mass function indicates a minimum mass for the secondary of 
$3.2 \pm 0.3$M$_{\odot}$, indicating intermediate or 
higher initial masses for the components. 

	A source of uncertainty in the light curve elements of PW Pup 
is the possible presence of light from an unresolved companion, and the high 
GAIA RUWE statistic provides evidence that such a body may be present. 
While the visible spectrum of PW Pup shows possible evidence
of light from a faint (when compared with the F
supergiant) B star \citep{hum1975}, the 
amplitude of the light curve cannot be reproduced if such a star is the 
primary in the eclipsing system. Nevertheless, 
the B star may still be the cause of the large RUWE statistic. 

	High angular resolution images of PW Pup and/or high resolution
spectra may confirm the presence of a hot companion star. The presence of a 
close third body can also be gauged using the light travel time effect 
\citep[][]{irw1952}, although this requires light curves spanning a 
broad range of epochs, making this a time (and 
resource) consuming process. Finally, high resolution 
spectroscopy, can be used to identify objects that are 
not moving in sync with the components. Objects that contribute light 
at less than the 10\% level might be detected \citep[e.g.,][]{menetal2023}.

	Aside from basic system elements, the light curves 
contain other clues about the evolutionary states of the 
systems. If mass is being transfered from the higher mass star to the 
lower mass companion then a high mass transfer rate is expected 
\citep[e.g.,][]{pac1971}. In many cases the impact of large scale mass 
transfer onto an accretion disk around the receiving star 
causes prominent photometric signatures, such as 
propogating bumps in the light curves. However, there are stream 
and disk geometries that may produce exceptions \citep[e.g.,][]{menetal2016}.
While there is no significant scatter in the SMC005609 $I$ light curve, 
cycle-to-cycle scatter at various phases is evident in the 
visible/red light curves of SMC004701 and PW Pup. 

	H$\alpha$ emission has also been detected from all 
three systems. This could originate from an accretion 
disk, although this is not a unique source of this 
emission. For example, H$\alpha$ emission originates from multiple 
sources in the $\beta$ Lyrae system \citep[e.g.,][]{broetal2021}. 
In summary, the spectrophotometric information indicates that the systems 
have almost certainly interacted, and are continuing to do so to this day.

	The nature of the primaries in these systems is of interest 
for assessing their future evolution. In traditional 
Algol systems the primary is usually unevolved, due to the transfer 
of angular momentum and mass from the secondary, both of which delay the 
evolution of the primary \citep[e.g.,][and references 
therein]{pus1996}. Signatures from a faint B star may be present in 
the spectra of SMC005609 (Section 1) and PW Pup \citep[][]{hum1975}. 
However, the light curves of these systems could not be reproduced with a 
faint B star primary, indicating that if a faint hot companion 
is present then it is not part of either eclipsing system.

	The primaries of the systems studied here are yellow giants, 
and this raises two possibilities. The first is that the primaries 
are the bloated remnants of what were the initially more massive 
stars that have lost much of their mass to the secondaries, 
which are now intermediate mass main sequence stars that 
are obscured by a circumstellar disk or shell. Each 
T$_2$ measurement made here is then the temperatures 
of the outer region of the shell; T$_2$ does not 
reflect the temperature of the hidden star. While highly evolved, 
the primaries are not red supergiants because their Roche surface limits 
their size. They have a blue color because a high effective 
temperature is required to shed their luminosity within the confines of their 
Roche lobes.

	An interesting second possibility is that the primaries may have been 
the initially lower mass star, but have received mass from the present day 
secondary and have now evolved off of the main sequence, but have not yet 
evolved into contact with their Roche surfaces. 
The long periods of the systems studied here 
make them possible progenitors of systems that will experience dynamically 
unstable mass loss \citep[e.g.,][]{geetal2020}. 
This is a potentially exciting stage of evolution, as such a system 
may form a common envelope when the primary fills its Roche lobe 
and sends a large mass flow to the cool secondary 
\citep[e.g., review by][]{ivaetal2013}. 

	We close by emphasizing that semi-detached models like those 
adopted here assume that the components in these systems are stars. 
Unfortunately, determining the nature of the secondary is problematic 
with the existing information. However, the ratios of luminosity and 
brightness listed in Table 2 suggest that the spectrum of a stellar secondary 
in SMC004701 may not prove difficult to detect at 
visible wavelengths, thereby allowing the nature of this object to 
be examined. For SMC005609 and PW Pup it is likely that 
the detection of lines from a stellar secondary will require observations 
at wavelengths $\geq 1\mu$m. 

	It is possible that the secondary may be 
a heavily obscured hot main sequence star that has received mass 
from the primary and is encased in a red shell. If 
an optically thick disk were present that fully engulfed the secondary 
then the photospheric signatures of that star may be 
invisible. The main body of the disk would 
likely appear as a flattened, luminous, cool source that would be 
faint at visible wavelengths except for emission lines where 
the disk is excited by, say, interactions with a mass stream. In fact,
the W1--W2 colors of SMC004701 vary with phase with an amplitude that is 
suggestive of a cool circumstellar component, although this behavior is not 
seen in the other systems. Pursuing opaque disk models may prove fruitful 
when a more complete body of spectroscopic information that could be used to 
probe disk signatures becomes available.

\acknowledgements{It is a pleasure to thank the anonymous referee for 
providing comments that greatly improved the paper. 
This work has made use of the WISE image 
(https://doi.org/10.26131/irsa151), 2MASS image 
(https://doi.org/10.26131/irsa121), and NASA Extragalactic Database services
(https://doi.org/10.26132/ned1), at the NASA/IPAC Infrared Science Archive 
which is funded by the National Aeronautics and Space Administration and
operated by the California Institute of Technology. This study has 
also used data from the European Space Agency (ESA) mission
{\it Gaia} (\url{https://www.cosmos.esa.int/gaia}), processed by the {\it Gaia}
Data Processing and Analysis Consortium (DPAC,
\url{https://www.cosmos.esa.int/web/gaia/dpac/consortium}). Funding for the DPAC
has been provided by national institutions, in particular the institutions
participating in the {\it Gaia} Multilateral Agreement.}

\end{document}